\documentstyle[12pt]{article}
\title{Indistinguishable particles and hidden variables}
\author{Adonai S. Sant'Anna \and D\'ecio Krause\\\\Dep. Matem\'atica\\Universidade Federal do Paran\'a\\C.P. 19081, Curitiba, PR, 80.530-900, Brasil}
\begin{document}
\maketitle
\begin{abstract}
We establish an axiomatic framework for indistinguishability of quantum particles in terms of hidden variables, which gives an ontology for microscopic particles. Such an axiomatic framework is set-theoretical. We also discuss the quantum distribution functions with the help of our axioms.
\end{abstract}

\newtheorem{definicao}{Definition}[section]
\newtheorem{teorema}{Theorem}[section]
\newtheorem{lema}{Lema}[section]
\newtheorem{corolario}{Corollary}[section]
\newtheorem{proposicao}{Proposition}[section]
\newtheorem{axioma}{Axiom}[section]
\newtheorem{observacao}{Observation}[section]

\section{Introduction}

	In classical physics it is possible to label individual particles, even in the case that they look alike. But in quantum mechanics, it is not possible, using the language of the physicist, to keep track of individual particles in order to distinguish `identical' particles. It is not possible to label electrons, for example, even in principle. The reason is that it is not possible to specify more than a complete set of commuting observables for each quantum particle. Yet, we cannot ``follow the trajectory because that would entail a position measurement at each instant of time, which necessarily disturbs the system'' \cite{Sakurai-94}.  We consider that this is true for a quantum theory with no ontological picture. We suggest in this paper a description for quantum mechanics that allows, in principle, to distinguish particles that are physically indistinguishable. There is no contradiction in our words, since we distinguish those particles at the ontological level.

	The search for axioms like those of set theories for dealing with collections of indistinguishable elementary particles was posed by Yu. Manin \cite{Manin-74}, in 1974, as one of the important problems of present day researches on the foundations of mathematics. As he said:

\begin{quote}
{\em I would like to point out that it is rather an extrapolation of common-place physics, whether we can distinguish things, count them, put them in some order, etc.. New quantum physics has shown us models of entities with quite different behaviour. Even {\em sets} of photons in a looking-glass box, or of electrons in a nickel piece are much less Cantorian than the {\em sets} of grains of sand.

	The twentieth century return to Middle Age scholastics taught us a lot about formalisms. Probably it is time to look outside again. Meaning is what really matters.}\cite{Manin-74}
\end{quote}

	Other authors \cite{Krause-92} \cite{Chiara-93} \cite{Krause-95} have considered that standard set theories are not adequate to represent microphysical phenomena as they are presented by the standard formulation of quantum mechanics. It is argued that the ontology of microphysics apparently does not reduce to that one of usual sets. In this paper we present a negative answer to this conjecture. We show that it is possible to give a set-theoretical framework for indistinguishability of quantum particles, specially for the ontology of quantum physics\footnote{In \cite{daCosta-96}, da Costa and Krause show that it is possible to establish set-theoretical models for quantum systems, since quasi-set theory can be translated into the usual Zermelo-Fraenkel set theory with the Axiom of Choice. Such a translation is related with the Heisenberg's paradox: ``The Copenhagen interpretation of quantum theory starts from a paradox. Any experiment in physics, whether it refers to the phenomena of daily life or to atomic events, is to be described in terms of classical physics.''}. Our main tool is the use of hidden variables. In this sense our solution to deal with physically indistinguishable particles is different from the approach proposed by Manin. We could consider this use of hidden variables as a try to complete the usual description of quantum particles. As van Fraassen remarks \cite{vanFraassen-85}:

\begin{quote}
{\em if two particles are of the same kind, and have the same state of motion, nothing in the quantum mechanical description distinguishes them.}
\end{quote}

	In this sense, quantum mechanics needs something more to distinguish particles, in order keep the classical mathematics to describe the theory. We propose that this something more could be hidden variables.

	The approach in terms of quasi-set theories to deal with indistinguishable objects is not appropriate to label quantum particles in order to obtain Bose-Einstein or Fermi-Dirac statistics if we are interested to follow the same mathematical techniques used by the physicist. In the hidden variables picture such a problem does not exist. We can easily label particles which are physically indistinguishable, since we assume that each particle has a different value to its hidden variable. Hence, with this approach it is possible to justify the quantum distribution functions as well as the symmetrical and antisymmetrical states of collections of quantum particles.

	It is well known the use of hidden variables in physics, specially in the description of quantum mechanics due to D. Bohm. Bohm \cite{Bohm-94} considered that the electron, e.g., ``has more properties than can be described in terms of the so-called `observables' of the quantum theory.'' He used hidden variables to give a deterministic picture to the ontology of quantum mechanics, although quantum systems behave in a probabilistic fashion, from the experimental point of view. Here, we preserve the concept of hidden variable as something that corresponds to inner properties of physical objects\footnote{We do not intend to discuss the concept of physical object. In the present text we consider that this concept is intuitivelly established.} that, at present, are not measured in laboratories. But our use of hidden variables is quite different, in principle, from that one of Bohm, since it has nothing to do with any explanation to the probabilistic behaviour of quantum phenomena.

	Our approach is out of the range of the proofs on the impossibility of hidden variables in the quantum theory, like von Neumann's theorem \cite{vonNeumann-55}, Gleason's work \cite{Gleason-57}, Kochen and Specker results \cite{Kochen-67} or Bell's inequalities \cite{Bell-87}. There are other works which claim to show that no distribution of hidden variables can account for the statistical predictions of the quantum theory. But in our ontological description of particles, specially quantum particles, we are not interested in the statistical aspects of quantum theory. Our concern is only with the so-called indistinguishability among particles.

	It is also well known that systems containing $n$ indistinguishable quantum particles are either totally symmetrical under the interchange of any pair (bosons) or are totally antisymmetrical (fermions). Our question is: if we have a system of $n$ indistinguishable particles, how can we label them in order to make the mentioned interchange of any pair? Usually it is said that we can mathematically label the particles. But if that is the case, we have an important mathematical concept that does not correspond to any physical interpretation: the label of physically indistinguishable particles. In the present paper we say that we can ontologically label each particle by the use of hidden variables which correspond to inner properties that are not characterized by the observables. This means that we can establish two kinds of identity: the physical and the ontological. Two particles physically indistinguishable (they have the same physical properties, in a sense to made clearer in the text) are always ontologically different or distinguishable. In other words, a system of $n$ quantum particles does never have two particles ontologically indistinguishable or even two particles with the same value for their respective hidden variables.

	Lowe \cite{Lowe-94} has suggested that quantum particles are genuinely (in a fundamentally ontological sense) {\em vague} objects. He considers a situation in which a free electron $a$ is captured by an atom to form a negative ion which then emits an electron labeled $b$ and notes that,

\begin{quote}
{\em according to currently accepted quantum mechanical principles there may be no objective fact of the matter as to whether or not $a$ is identical with $b$. It should be emphasized that what is being proposed here is not merely that we have no way of telling whether or not $a$ and $b$ are identical, which would imply only an epistemic indeterminacy. It is well known that the sort of indeterminacy pressuposed by orthodox interpretations of quantum theory is more than merely epistemic - it is ontic.}
\end{quote}

	According to our ontological picture, each electron has a well defined hidden variable, which allows to attribute a label. But in our axiomatic treatment we are not able to describe the dynamics of the process remarked by Lowe. We only know that if electron $b$ has the same hidden variable of electron $a$, then they are identical in the sense that they are the same particle. But if $a$ and $b$ have different values for their hidden variables, then we are really talking about {\em two} electrons. In this case, they are two indistinguishable particles, but they still are two electrons (ontologically distinguishable).

	Dalla Chiara \cite{Chiara-85} develops a quantum logical semantics for identical particles in which proper names and definite descriptions may lack a precise {\em denotatum} within some possible worlds. In \cite{Chiara-93} Dalla Chiara and Toraldo di Francia conclude, as a philosophical consequence of this semantics, that there is no {\em trans-world} identity. But it is obvious that this inexistence of a trans-world identity is a consequence of the hypothesis that there is no trans-world identity. The semantics developed by Dalla-Chiara comes from the observation that the world of {\em identical particles} in microphysics gives rise to examples of uncertain and ambiguous denotation relations. It is clear that Dalla-Chiara did not consider the possibility of ontological denotation relations.

	In the next section we present an axiomatic framework for ontologically distinguishable particles in terms of a set-theoretical predicate. This predicate allows to cope with collections of physically indistinguishable particles as sets. Then, in section 3 we present the physical consequences of this picture, with special attention to the quantum distribution functions.

\section{Set-Theoretical Predicate for Ontologically Distinguishable Particles}

	We are not interested to give an axiomatic framework for quantum physics, quantum mechanics or even mechanics. Our concern is with the process of labeling indistinguishable particles, so widely used by physicists.

	Our system has seven primitive notions: $\lambda$, $X$, $P$, $m$, $M$, $\equiv$, and $\doteq$. $\lambda$ is a function $\lambda:N\rightarrow${\bf R}, where $N$ is the set $\{1,2,3...,n\}$, $n$ is a nonnegative integer, and {\bf R} is the set of real numbers; $X$, and $P$ are finite sets; $m$ and $M$ are predicates defined on elements of $P$; and $\equiv$ and $\doteq$ are binary relations between elements of $P$. Intuitivelly, the images $\lambda_{i}$ of the function $\lambda$, where $i\in N$, correspond to the so-called hidden variables. $X$ is to be interpreted as one set such that each one of its elements corresponds to measurements of physicall observables of one particle. Such measurements can be precisely characterized by the {\em generalized operational definition of a physical quantity}\footnote{Although this definition for physical quantity receives some criticisms by science philosophers, we consider, as Dalla-Chiara and Toraldo di Francia \cite{Chiara-79}, that such a definition reflects a methodology that is largely accepted by physicists.}. Basically, a physical quantity is defined by a union $C = \bigcup\{ C_{k}\}$ over a set $\{ C_{k}\}$ of equivalence classes of measuring procedures, such that the set $\{ C_{k}\}$ is connected and each $C_{k}$ is defined over a well-determined class $\Sigma_{k}$ of physical systems, where $\Sigma_{k}\neq\Sigma_{l}$ for $k\neq l$. For details see \cite{Chiara-79}. The elements of $X$ are denoted by $x$, $y$, etc. $P$ is to be physically interpreted as the set of particles. $m(p)$, where $p\in P$, means that $p$ is a microscopic particle. $M(p)$ means that $p\in P$ is a macroscopic particle. $\equiv$ corresponds to the ontological identity between particles and $\doteq$ corresponds to the physical identity between particles.

\begin{definicao}
$\Lambda$ is the set of images of the function $\lambda$.
\end{definicao}

\begin{definicao}
${\cal D_{O}} = \langle\lambda,X,P,m,M,\equiv,\doteq\rangle$ is a system of ontologically distinguishable particles if and only if the following axioms are satisfied:
\begin{description}
\item [D1] $\lambda:N\rightarrow {\bf R}$ is an injective function.
\item [D2] $P\subset X\times\Lambda$.
\item [D3] $x\neq y\rightarrow\neg (\langle x,\lambda_{i}\rangle\in P \wedge \langle y,\lambda_{i}\rangle\in P)$
\item [D4] $\langle x,\lambda_{i}\rangle\equiv \langle y,\lambda_{j}\rangle \leftrightarrow x=y\wedge i=j$.
\item [D5] $\langle x,\lambda_{i}\rangle\doteq \langle y,\lambda_{j}\rangle\leftrightarrow x=y$.
\item [D6] $\langle x,\lambda_{i}\rangle\doteq \langle y,\lambda_{j}\rangle\rightarrow m(\langle x,\lambda_{i}\rangle)\wedge m(\langle y,\lambda_{j}\rangle)$.
\item [D7] $m(\langle x,\lambda_{i}\rangle)\vee M(\langle x,\lambda_{i}\rangle)\rightarrow\neg(m(\langle x,\lambda_{i}\rangle)\wedge M(\langle x,\lambda_{i}\rangle)$.
\end{description}
\end{definicao}

	Axiom {\bf D1} corresponds to say that the cardinality of $\Lambda$ coincides with the cardinality of $N$ ($\#\Lambda = \# N$). Axiom {\bf D2} just says that particles are represented by ordered pairs\footnote{In \cite{daCosta-94} da Costa and Krause discuss the possible representation of a quantum particle in terms of an ordered pair $\left< E,L\right>$, where $E$ corresponds to a predicate which in some way characterizes the particle in terms, e.g., of its rest mass, its charge, and so on. $L$ denotes an apropriate label, which could be, for example, the spatio-temporal location of the particle. Even in the case that the particles (in a system) have the same $E$, they might be distinguished by their labels. In this case, we are dealing with a classical representation of the particles. But if the particles have the same label, the tools of classical mathematics cannot be applied. In our picture, according to axioms {\bf D1}-{\bf D3}, it is prohibited a system where two particles have the same (ontological) label.}, where the first element corresponds to the physical properties measurable in laboratory, and the second element corresponds to the hidden inner property that allows to distinguish particles at the ontological level. Yet, axioms {\bf D2} and {\bf D3} guarantee that $\# P=\# N=\#\Lambda$, which corresponds to the number of particles of the system. In other words, two particles in a system of ontologically distinguishable particles do never have the same hidden variable. Axiom {\bf D4} says that two particles are ontologically indistinguishable if and only if they have the same physical properties and the same hidden variables. Axiom {\bf D5} means that two particles are physically indistinguishable if and only if they have the same physical properties. Axiom {\bf D6} corresponds to say that if two particles are physically indistinguishable, then both of them are microscopic or quantum particles. Axiom {\bf D7} means that one particle cannot be microscopic and macroscopic.

	One could argue that function $\lambda$ is desnecessary, since we could interpret the elements of $N$ as the hidden variables that allow to label particles even when they are physically indistinguishable. We consider that this is not a satisfactory assumption, since we are interested to emphasize that the hidden variables correspond to inner properties of all particles, macroscopic or microscopic, that are not measurable in laboratory, at least in the present. Our hidden variables are not just a mathematical tool to label particles. We mean that it is possible that some day, some experimental physicist discovers a new physical property of quantum particles that allows to label them. Such a physical observable would correspond to our hidden variables. To interpret the images of $\lambda$ as the hidden variables means that the measurements of this possible future observable would assume values in the set of real numbers. Obviously, our concept of hidden variable could be extended to a function $\lambda:N\rightarrow V$, where $V$ is a vector space.

	The theorem given below says that two macroscopic particles cannot be physically indistinguishable, or, in other words, we can always label macroscopic particles in one laboratory.

\begin{teorema}
$M(\langle x,\lambda_{i}\rangle)\wedge M(\langle y,\lambda_{j}\rangle)\rightarrow \neg(\langle x,\lambda_{i}\rangle\doteq \langle y,\lambda_{j}\rangle)$.\label{macro}
\end{teorema}
{\bf Proof}: If $M(\langle x,\lambda_{i}\rangle)\wedge M(\langle y,\lambda_{j}\rangle)$, then, by axiom {\bf D7},\\ $\neg(m(\langle x,\lambda_{i}\rangle)\wedge m(\langle y,\lambda_{j}\rangle)$. Hence, by axiom {\bf D6},\\$\neg(\langle x,\lambda_{i}\rangle\doteq \langle y,\lambda_{j}\rangle)$.$\Box$\\

	The theorem given below is relevant for the discussions about quantum distribution functions in the next section.

\begin{teorema}
If $X$ is a unitary set and $\# N\geq 2$, then the system of ontologically distinguishable particles has only microscopic particles.\label{unitaryX}
\end{teorema}
{\bf Proof}: If $\# N\geq 2$, then $\# P\geq 2$, by axioms {\bf D1}, {\bf D2}, and {\bf D3}. This means that we have a system with more than just one particle. But all these particles have the same physical properties, since we assume, by hypothesis, that $X$ is unitary. Hence, all particles are physically indistinguishable, by axiom {\bf D5}. So, all particles are microscopic, by axiom {\bf D6}.$\Box$

\section{Distribution Functions for Quantum Particles}

	Our main objective in this section is to show how to establish the sufficient conditions to obtain the quantum distribution functions in our picture for indistinguishable particles in terms of hidden variables.

	To obtain the quantum distribution functions in the standard way it is necessary to assume that the quantum particles are indistinguishable. In the case of fermions, we assume also the {\em Pauli exclusion principle}. Bosons do not satisfy such a principle. But the fundamental assumption of indistinguishability between quantum particles means that either we deal with this collection of particles as a quasi-set or we assume the existence of hidden variables. The second alternative allows to deal with collections of physically indistinguishable particles as sets. In this section we present an interpretation of Bose-Einstein and Fermi-Dirac statistics in set-theoretical terms.

	The Pauli exclusion principle states that two or more fermions cannot occupy the same state. This occurs because a state like $\mid k'\rangle\mid k'\rangle$ is necessarily symmetrical, which is not possible for a fermion. But different states cannot be used to label fermions, since a fermion can change its state. In the case of bosons, the situation is more dramatic, since we can have several bosons occupying the same single state. If we have a collection of indistinguishable bosons or indistinguishable fermions, is this collection a set? In our picture the answer is positive.

	The fermion case will be discussed first. To cope with a collection of fermions we consider, as a first assumption, an ${\cal D_{O}}$-system with an unitary set $X$. We know that if $X$ is unitary in a system with more than one particle, then all particles are microscopic, according to theorem \ref{unitaryX}. So, fermions are microscopic particles because they are physically indistinguishable. It must be emphasized that to deal with fermions, we consider that the unique element $x$ of $X$ corresponds to the measurements of a complete set of commuting observables, otherwise it would be impossible to satisfy axiom {\bf D5}, since non-commuting observables do satisfy Heisenberg principle of uncertainty. Our second assumption is the Pauli exclusion principle written in terms of our language. But before that, we need to establish the meaning of symmetrical and antisymmetrical states.

	For the sake of simplicity, we consider a system of two physically indistinguishable particles, ontologically labeled particle $\lambda_{1}$ and particle $\lambda_{2}$. Suppose that, in the Hilbert space formalism, particle $\lambda_{1}$ is characterized by the state vector $\mid k'_{\lambda_{1}}\rangle$, where $k'$ corresponds to a collective index for a complete set of observables (commuting or not), or, in other words, $k'$ contains more physical information in terms of observables than $x$. Actually, if we were concerned with a rigorous notation, we should denote the state of particle $\lambda_{1}$ as $\mid k'-x,\langle x,\lambda_{1}\rangle\rangle$, where $k'-x$ corresponds to the extra physical information that is not available in $x$. But, in practice, we are abbreviating the notation. Likewise, we denote the ket of the remaining particle $\mid k''_{\lambda_{2}}\rangle$. The state ket for the two particles system is

\begin{equation}
\mid k'_{\lambda_{1}}\rangle\mid k''_{\lambda_{2}}\rangle.
\end{equation}

	If a measurement is performed on this system, it may be obtained $k'$ for one particle and $k''$ for the other one. But, in the laboratory, it is not possible to know if the state ket of the system is $\mid k'_{\lambda_{1}}\rangle\mid k''_{\lambda_{2}}\rangle$, $\mid k''_{\lambda_{1}}\rangle\mid k'_{\lambda_{2}}\rangle$ or any linear combination $c_{1}\mid k'_{\lambda_{1}}\rangle\mid k''_{\lambda_{2}}\rangle + c_{2}\mid k''_{\lambda_{1}}\rangle\mid k'_{\lambda_{2}}\rangle$. This is called the exchange degeneracy, which means that to determine the eigenvalue of a complete set of observables does not uniquely specify the state ket.

	Using a notation similar to Sakurai's \cite{Sakurai-94} we define the permutation operator $P_{12}$ by

\begin{equation}
P_{12}\mid k'_{\lambda_{1}}\rangle\mid k''_{\lambda_{2}}\rangle = \mid k''_{\lambda_{1}}\rangle\mid k'_{\lambda_{2}}\rangle.\label{interchange}
\end{equation}

	It is obvious that $P_{21} = P_{12}$ and $P_{12}^{2} = 1$. In the case we are discussing:

\begin{equation}
P_{12}\mid k'_{\lambda_{1}}\rangle\mid k''_{\lambda_{2}}\rangle = -\mid k'_{\lambda_{1}}\rangle\mid k''_{\lambda_{2}}\rangle,
\end{equation}
or, in the more general situation:
\[P_{ij}\mid\mbox{$n$ physically indistinguishable fermions}\rangle =\]
\begin{equation}
-\mid\mbox{$n$ physically indistinguishable fermions}\rangle,\label{fermions}
\end{equation}
where $P_{ij}$ is the permutation operator that interchanges the particle ontologically labeled as $\lambda_{i}$ and the particle ontologically labeled as $\lambda_{j}$, with $i$ and $j$ arbitrary but distinct elements of $N$. We must recall again that in equation (\ref{fermions}) the sentence ``$n$ physically indistinguishable fermions'' means that each arbitrary pair of fermions has the same values for measurements of a complete set of commuting observables.

	In our picture it is possible to count fermions, since we can label them and, so, to deal with collections of fermions as sets. These sets could be called ``ontological sets''. It is clear also what means to say that a system of fermions is totally antisymmetrical under the interchange of any pair, since now it is clear the meaning of the word ``interchange'' according to equation (\ref{interchange}). With this in mind we observe that, by equation (\ref{interchange}), 
\begin{equation}
P_{12}\mid k'_{\lambda_{1}}\rangle\mid k'_{\lambda_{2}}\rangle = \mid k'_{\lambda_{1}}\rangle\mid k'_{\lambda_{2}}\rangle,
\end{equation}
which contradicts equation (\ref{fermions}). Hence, as expected, fermions cannot occupy the same physical state, which is a translation of the exclusion principle in our language of hidden variables.

	The discussion about bosons is very similar and we let this case as an exercise for the reader.

	Since we characterized the permutation operator, symmetrical and antisymmetrical states, Pauli exclusion principle and the labeling of quantum particles, now we can easily deduce the quantum distribution functions by standard ways. For details see, for example, \cite{Garrod-95}.

	In texts like \cite{Sakurai-94} other physical consequences of the indistinguishability among quantum particles are cited. But all these effects are consequences of the symmetrical or antisymmetrical properties of quantum particles, which we have ever discussed.

\end{document}